\documentclass[12pt,a4paper]{article}
\usepackage[final]{showkeys}
\usepackage{amsmath}
\usepackage{amssymb}
\usepackage{amsthm}
\newcommand{\Eins}{\boldsymbol 1}

\newcommand{\g}[1]{\mathfrak{#1}}
\renewcommand{\r}[1]{\mathrm{#1}}
\renewcommand{\t}[1]{\texttt{#1}}
\newcommand{\co}[1]{\textsf{#1}}
\newcommand{\tr}{\mathop\mathrm{Tr}\nolimits}
\newtheorem*{lemm}{Lemma}
\newtheorem*{theo}{Theorem}

\title{Continuity of quantum conditional information}
\author{R.~Alicki~\footnote{Electronic address: 
\t{fizra@univ.gda.pl}} \\
{\normalsize Institute of Theoretical Physics and Astrophysics} \\
{\normalsize University of Gda\'nsk, Poland} 
\and
M.~Fannes~\footnote{Electronic address: 
\t{mark.fannes@fys.kuleuven.ac.be}} \\
{\normalsize Instituut voor Theoretische Fysica} \\
{\normalsize K.U. Leuven, Belgium} }

\date{}

\begin{document}

\maketitle

\begin{abstract}
We prove continuity of quantum conditional information $S(\rho^{12}|\,\rho^2)$ with respect to the uniform convergence of states and obtain a bound which is independent of the dimension of the second party. This can, e.g., be used to prove the continuity of squashed entanglement.
\end{abstract}

A, generally mixed, state of a bipartite system is given by a density matrix $\rho^{12}$ on a Hilbert space $\g H^{12} = \g H^1 \otimes \g H^2$. We shall, in order to avoid technical complications, restrict our attention to finite dimensional systems and not distinguish between the density matrix $\rho^{12}$ and its associated expectation functional
\begin{equation*}
a \mapsto \rho^{12}(a)  := \tr \rho^{12}\, a, \quad a \text{ a linear operator on } \g H^{12}.
\end{equation*}   
The restrictions of $\rho^{12}$ to the subsystems 1 and 2 are denoted by $\rho^1$ and $\rho^2$, e.g.\ 
\begin{equation*}
\rho^1(a) := \rho^1(a\otimes \Eins) = \tr \rho^{12}\, a\otimes \Eins, \quad a \text{ a linear operator on } \g H^1.
\end{equation*} 
The \co{von~Neumann entropy} $S(\rho)$ of a state $\rho$ is the quantity $\tr \eta(\rho)$ with 
$\eta(x):= -x\log x$ for $0<x\le1$ and $\eta(0)=0$. The \co{conditional information} $S(\rho^{12}|\,\rho^2)$ of $\rho^{12}$ with respect to the second system is the quantity
\begin{equation*}
S(\rho^{12}|\,\rho^2) := S(\rho^{12}) - S(\rho^2),
\end{equation*} 
$S(\rho^{12}|\,\rho^2)$ is also called \co{conditional entropy}. Finally, we need the uniform distance between states: 
\begin{equation*}
\|\rho-\sigma\|_1 := \sup_{a,\, \|a\|\le1} |\rho(a)-\sigma(a)| = \tr |\rho-\sigma|.  
\end{equation*}
In this last equation $|a|$ denotes the absolute value of a matrix (or an operator on Hilbert space). It is given by $|a|:=\sqrt{a^*a}$. For the matrix case, the eigenvalues of $|a|$ are often called the singular values of $a$.

\begin{theo}
\label{th1}
Take any two states $\rho^{12}$ and $\sigma^{12}$ on $\g H^{12} = \g H^1 \otimes \g H^2$ such that $\epsilon := \|\rho^{12}-\sigma^{12}\| < 1$ and let $d_1$ be the dimension of $\g H^1$, then the following estimate holds
\begin{equation}
\bigl| S(\rho^{12}|\,\rho^2) - S(\sigma^{12}|\,\sigma^2)\bigr| \le 4\epsilon \log d_1 + 2\eta(1-\epsilon) + 2\eta(\epsilon).
\label{th1.1} 
\end{equation}
In particular, the right-hand side of~(\ref{th1.1}) does not explicitly depend on the dimension of $\g H^2$.  
\end{theo}

\begin{proof}
The basic step in the proof is the introduction of an auxiliary state  
\begin{equation*}
\gamma^{12} := (1-\epsilon)\rho^{12} + |\rho^{12}-\sigma^{12}|.
\end{equation*}
As $0\le\epsilon\le1$, $\gamma^{12}$ is indeed a state. Next, dealing only with the non-trivial case $\epsilon>0$, we introduce two new states
\begin{equation*}
\tilde\rho^{12} 
:= \frac{1}{\epsilon}\, |\rho^{12}-\sigma^{12}|
\qquad\text{and}\qquad
\tilde\sigma^{12} 
:=  \frac{1-\epsilon}{\epsilon}\, (\rho^{12}-\sigma^{12}) 
+ \frac{1}{\epsilon}\, |\rho^{12}-\sigma^{12}|.
\end{equation*} 
A direct computation shows that
\begin{equation*}
 \gamma^{12} = (1-\epsilon)\, \rho^{12} 
 + \epsilon\, \tilde\rho^{12} 
 = (1-\epsilon)\, \sigma^{12} + \epsilon\, \tilde\sigma^{12}.
\end{equation*}
The situation precisely matches that of the theorem of 
Thales of Milete in planar geometry. 

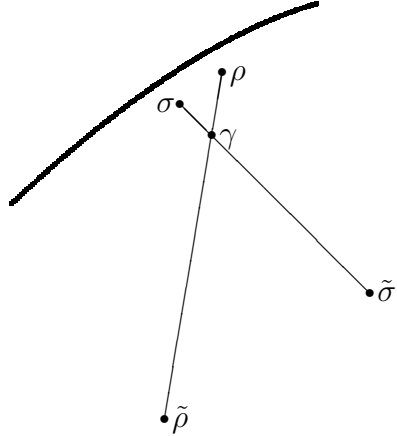
\begin{center}
\setlength{\unitlength}{7cm}
\begin{picture}(1,1)
\put(.34,.19){\line(1,6){.11}}
\put(.34,.19){\circle*{.015}}
\put(.355,.175){$\tilde\rho$}
\put(.43,.73){\circle*{.015}}
\put(.445,.715){$\gamma$}
\put(.45,.85){\circle*{.015}}
\put(.465,.835){$\rho$}
\put(.73,.43){\line(-1,1){.36}}
\put(.73,.43){\circle*{.015}}
\put(.745,.415){$\tilde\sigma$}
\put(.37,.79){\circle*{.015}}
\put(.325,.775){$\sigma$}
\linethickness{.5mm}
\qbezier(.05,.6)(.38,.91)(.63,.98)
\end{picture}
\end{center}

We now estimate
\begin{align*}
 &\bigl| S(\rho^{12}|\,\rho^2) - S(\sigma^{12}|\,\sigma^2)\bigr| \\ 
 &\quad\le \bigl| S(\rho^{12}|\,\rho^2) - 
 S(\gamma^{12}|\,\gamma^2)\bigr| 
 + \bigl| S(\sigma^{12}|\,\sigma^2) - S(\gamma^{12}|\,\gamma^2)\bigr| \\
 &\quad\le 4\epsilon \log d_1 + 2\eta(1-\epsilon) + 2\eta(\epsilon).
\end{align*}
The last inequality follows from the lemma. 
\end{proof}

\begin{lemm}
\label{le1}
Let $\rho^{12}$ and $\tilde\rho^{12}$ be two states on 
$\g H^{12}$, let $0\le\epsilon\le1$ and put 
$\gamma^{12} := (1-\epsilon)\, \rho^{12} 
+ \epsilon\, \tilde\rho^{12}$. 
If $d_1$ is the dimension of $\g H_1$, then
\begin{equation*}
 \bigl| S(\rho^{12}|\,\rho^2) - S(\gamma^{12}|\,\gamma^2)\bigr| 
 \le 2\epsilon \log d_1 + \eta(1-\epsilon) + \eta(\epsilon).
\end{equation*}
\end{lemm}

\begin{proof}
As the conditional entropy is concave,see~\cite{T}, we have
\begin{equation*}
 S(\gamma^{12}|\,\gamma^2) 
 \ge (1-\epsilon)\, S(\rho^{12}|\,\rho^2) 
 + \epsilon\, S(\tilde\rho^{12}|\,\tilde\rho^2).
\end{equation*}
Therefore
\begin{align}
 S(\rho^{12}|\,\rho^2) - S(\gamma^{12}|\,\gamma^2) 
 &\le \epsilon \Bigl( S(\rho^{12}|\,\rho^2) -
 S(\tilde\rho^{12}|\,\tilde\rho^2) \Bigr) 
\nonumber \\
 &\le 2\epsilon \log d_1. 
\label{le1.1}
\end{align}
Next, we use the concavity of the entropy
\begin{equation*}
 S(\gamma^2) \ge (1-\epsilon)\, S(\rho^2) 
 + \epsilon\, S(\tilde\rho^2)
\end{equation*}
and the upper bound
\begin{equation*}
 S(\gamma^{12}) \le (1-\epsilon)\, S(\rho^{12}) 
 + \epsilon\, S(\tilde\rho^{12}) + \eta(1-\epsilon) 
 + \eta(\epsilon)
\end{equation*}
to obtain
\begin{align*}
 S(\gamma^{12}|\,\gamma^2) 
 &= S(\gamma^{12}) - S(\gamma^2) \\
 &\le (1-\epsilon) \bigl( S(\rho^{12}) - S(\rho^2) \bigr) 
 + \epsilon \bigl( S(\tilde\rho^{12}) - S(\tilde\rho^2) \bigr) 
 + \eta(1-\epsilon) + \eta(\epsilon).
\end{align*}
Therefore
\begin{align}
 S(\rho^{12}|\,\rho^2) - S(\gamma^{12}|\,\gamma^2) 
 &\ge \epsilon \Bigl( S(\rho^{12}|\,\rho^2) -
 S(\tilde\rho^{12}|\,\tilde\rho^2) \Bigr) 
 - \eta(1-\epsilon) - \eta(\epsilon) 
\nonumber \\
 &\ge -2\epsilon \log d_1 - \eta(1-\epsilon) 
 - \eta(\epsilon). 
\label{le1.2}
\end{align}
Combining~(\ref{le1.1}) and~(\ref{le1.2}), the lemma follows.  
\end{proof}

M.~Horodecki~\cite{H} kindly explained us the relevance of 
continuity of conditional quantum information for obtaining 
the asymptotic continuity of the newly introduced 
\co{squashed entanglement} for mixed bipartite states. 
This quantity is the smallest conditional mutual information 
computed over the set of all finite dimensional extensions of the state:   
\begin{equation*}
 E_{\r{sq}(\rho^{12})} := \inf_{\rho^{123}}\ \frac{1}{2}\, 
 \Bigl(  S(\rho^{13}|\,\rho^3) - S(\rho^{123}|\,\rho^{23}) \Bigr).
\end{equation*}
It was introduced in~\cite{Tu} and shown to vanish on separable states 
and to be additive, see~\cite{CW}. Continuity in the state, however, remained an open issue, the missing piece being precisely our theorem.

As conditional entropy is a rather basic quantity in quantum 
information theory, it's continuity might prove useful 
in another context. We would also like to point out that the 
theorem provides a concise proof, though with non-optimal constants, 
of the continuity of von~Neumann entropy
\begin{equation*}
 |S(\rho) - S(\sigma)| \le 2 \epsilon\log d + 2 \eta(\epsilon) 
 + 2 \eta(1-\epsilon)
\end{equation*} 
with $\epsilon:=\|\rho-\sigma\|_1$ and $d$ the dimension of the 
space spanned by $\rho$ and $\sigma$.
\medskip
   
\noindent
\textbf{Acknowledgements:} 
This work was done while one of us (M.F.) visited the Institute of 
Theoretical Physics and Astrophysics in Gda\'nsk. It is a pleasure 
to acknowledge the fruitful atmosphere and warm hospitality during 
this visit. This work was partially supported by KBN grant 2P03B08425 
and by EC grant RESQ IST-2001-37559.

\end{document}